\documentclass{appolb}
\usepackage{graphicx}

\begin{document}
\title{Recent progress in the studies of neutron rich and high-$Z$ systems
within the covariant density functional theory%
\thanks{Presented at XXIII Nuclear Physics Workshop ``Marie \&  Pierre Curie'', 
Kazimierz Dolny, Poland 2016}%
}
\author{A.~ V.~ Afanasjev, S.~ E.~ Agbemava, D.~ Ray
\address{Department of Physics and Astronomy, Mississippi State
University, MS 39762, USA}
\\
\vspace{0.5cm}
{P.~ Ring}
\address{Fakult\"at f\"ur Physik, Technische Universit\"at M\"unchen,
 D-85748 Garching, Germany}
}
\maketitle
\begin{abstract}
   The analysis of statistical and systematic uncertainties and their
propogation to nuclear extremes has been performed. Two extremes of
nuclear landscape (neutron-rich nuclei and superheavy nuclei) have been
investigated. For the first extreme, we focus on the ground state properties.
For the second extreme, we pay a particular attention to theoretical
uncertainties in the description of fission barriers of superheavy nuclei
and their evolution on going to neutron-rich nuclei.
\end{abstract}

\section{Introduction}

\begin{figure*}[ht]
\centering
\includegraphics[angle=0,width=12.0cm]{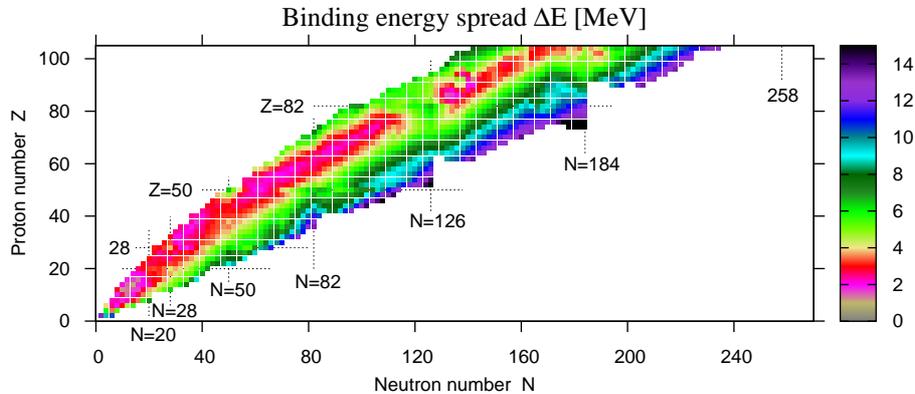}
\caption{The binding energy spread $\Delta E(Z,N)$ as a
function of proton and neutron number.
$\Delta E(Z,N)=|E_{\rm max}(Z,N)-E_{\rm min}(Z,N)|$, where
$E_{\rm max}(Z,N)$ and $E_{\rm min}(Z,N)$ are the largest and
the smallest binding energies for each ($N,Z$)-nucleus
obtained with the four covariant energy density functionals 
NL3*, DD-PC1, DD-ME2 and DD-ME$\delta$. From Ref.\ 
\cite{AARR.14}.}
\label{DE_spread_total}
\end{figure*}

  The physics of neutron rich (up to the neutron drip line) and extreme $Z$ superheavy
nuclei and the question of the reliability of theoretical extrapolations to such
systems are of paramount importance considering the construction of next generation
facilities such as FRIB, FAIR, Superheavy Elements Factory, etc. which will be
operational in the beginning of next decade. Even with these facilities the expansion
of the experimentally known nuclear landscape will be modest (see Fig.\ \ref{FRIB-impact}
below) and still a huge number of nuclei will be beyond of experimental reach. However,
these nuclei are important in nuclear astrophysical processes such as the r-process
\cite{MSMA.16} and fission recycling in neutron star mergers \cite{GBJ.11}.

 Thus the quality of an extrapolation of model predictions to unknown regions
of the periodic chart is an important issue.  This quality is characterized 
by systematic and statistical uncertainties \cite{DNR.14}. Statistical uncertainties emerge from
the details of the fitting protocol such as the choice of experimental data
and the selection of adopted errors; they characterize a given functional. Systematic
uncertainties emerge from the underlying theoretical approximations; they characterize
a selected group  of  functionals. In nuclear density functional theories (DFT), there are
several major sources of approximations, as for instance the general form of the functional, 
the range of the effective interaction or the form of its density dependence. In presently used
covariant density functional theory (CDFT) \cite{VALR.05} the density dependence is introduced 
either through an explicit dependence of the coupling constants \cite{TW.99,DD-ME2,DD-PC1} or via
non-linear meson couplings \cite{BB.77,NL3*}. Point coupling and meson exchange
models have an interaction of zero and of finite range, respectively
\cite{VALR.05,DD-ME2,DD-PC1,NL3*}.  As a consequence, at present several major classes of
covariant energy density functionals (CEDF) exist dependent on the combination
of above mentioned features (see Ref.\ \cite{AARR.14} for detail).

  In recent years a number of comprehensive investigations of systematic uncertainties
in the ground state observables and their propagation with particle numbers have been
performed by us using the NL3* \cite{NL3*}, DD-ME2 \cite{DD-ME2}, DD-ME$\delta$
\cite{DD-MEd}, DD-PC1 \cite{DD-PC1} and PC-PK1 \cite{PC-PK1} CEDFs as state-of-the-art
representatives of above mentioned major classes of CEDFs. These studies cover the properties
and related systematic uncertainties of different physical observables for all even-even
nuclei with $Z\leq 106$ \cite{AARR.14,AARR.13},  for the position of the two-neutron drip line
\cite{AARR.14,AARR.13,AA.16}, for octupole deformed \cite{AAR.16} and superheavy
nuclei \cite{AANR.15}.  Note that the analysis of theoretical uncertainties has
also been performed in Skyrme DFTs in Refs.\ \cite{Eet.12,KENBGO.13,GDKTT.13,MSHSWN.15}, but
the focus was mostly on the statistical uncertainties.   In this manuscript we deal
with covariant energy density functionals. In Sect.\ \ref{NR} we investigate
the statistical uncertainties in the description of the ground state properties of spherical
nuclei and their relation to systematic ones. Sect. \ref{FB} presents a study of statistical
and systematic uncertainties in the description of inner fission barriers in superheavy
nuclei. Finally, Sect.\ \ref{concl} summarizes our conclusions.

\section{Ground state properties of neutron-rich nuclei}
\label{NR}

\begin{figure*}[ht]
\centering
\includegraphics[angle=-90,width=10.0cm]{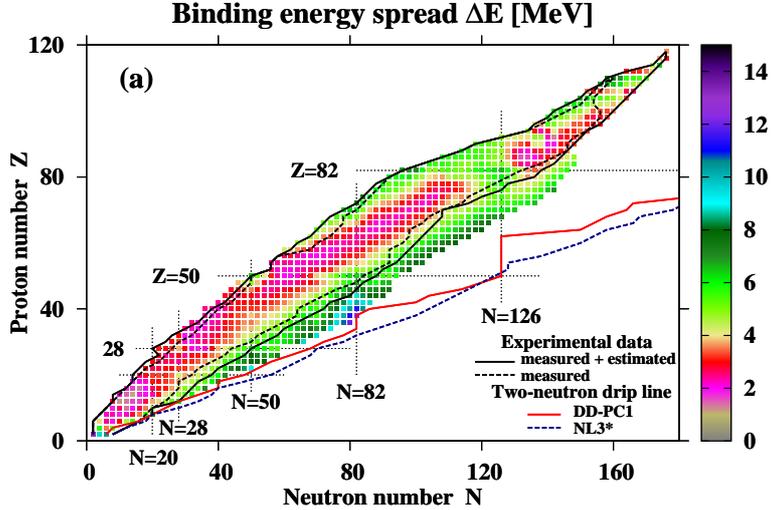}
\caption{The impact of future measurements on the nuclear landscape.
The squares show the results presented in Fig.\ \ref{DE_spread_total}
but only for the nuclei which are currently known and which will be
measured with FRIB. The regions of the nuclei with measured and
measured+estimated masses are enclosed by dashed and solid black
lines, respectively. The squares beyond these regions indicate the
nuclei which may be measured with FRIB. The line formed by the most
neutron-rich nucleus in each isotope chain accessible with FRIB is
called as ``FRIB limit''.
The same colormap as in Fig.\ \ref{DE_spread_total} is used here, but
the ranges of particle numbers for the vertical and horizontal axis are
different from the ones in Fig.\ \ref{DE_spread_total}. The two-neutron
drip lines are shown for the CEDFs NL3* and DD-PC1 by blue dashed and
solid red lines, respectively. From Ref.\ \cite{AA.16}.
}
\label{FRIB-impact}
\end{figure*}

   Systematic theoretical uncertainties in the prediction of
binding energies for the four CEDFs NL3*, DD-ME2, DD-PC1
and DD-ME$\delta$ are shown in Fig.\ \ref{DE_spread_total}.
While the spreads in the predictions of binding energies stay
within 5-6 MeV in the region of the known nuclei \cite{AARR.14,AA.16}
(see also the region enveloped by solid black line in Fig.\
\ref{FRIB-impact}), they increase drastically when approaching
the  neutron-drip line where they can reach 15 MeV. 
This is a consequence of the poorly defined isovector properties of 
the existing CEDFs.

\begin{figure*}[ht]
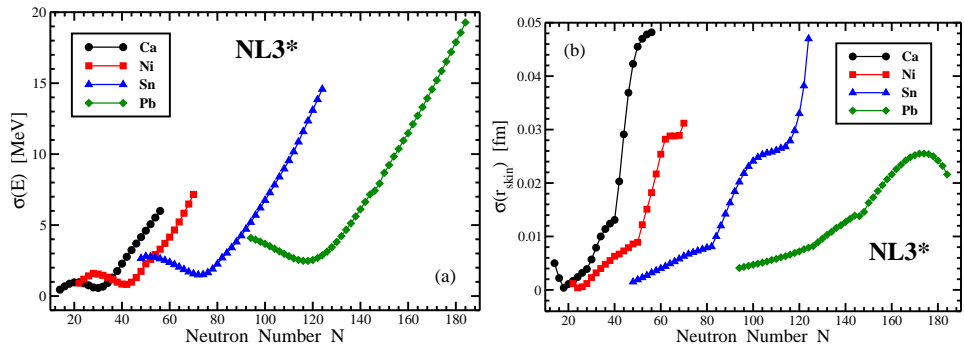

\centering
\includegraphics[angle=0,width=6.2cm]{fig-2-a.eps}
\includegraphics[angle=0,width=6.2cm]{fig-2-b.eps}
\caption{Statistical uncertainties in binding energies (panel
(a)) and neutron skins (panel (b)). All even-even nuclei
between the two-proton and two-neutron drip lines are included.}
\label{ground-stat-uncert}
\end{figure*}

  Statistical theoretical uncertainties in binding energies and
neutron skins are shown in Fig.\ \ref{ground-stat-uncert}. These
uncertainties are expressed as standard deviations  $\sigma (E)$
and $\sigma (r_{\rm skin})$ for a set of ``reasonable'' variations
of the original functional defined according to Ref.\ \cite{DNR.14}.
They are calculated in the spherical relativistic Hartree-Bogoliubov
(RHB) framework with the CEDF NL3* for the Ca, Ni, Sn and Pb isotope 
chains
from the two-proton to the two-neutron drip line.
The $\sigma (E)$ values are close to the adopted errors of the fitting
protocol for the nuclei used in the fitting. However,
they rapidly increase with increasing neutron number so that for
the nuclei in the vicinity of the two-neutron drip line they reach
values comparable with the spreads in binding energies shown
in Fig.\ \ref{DE_spread_total}. This fact should not be used as an
argument in favor of the similarity of statistical and systematic
uncertainties for binding energies since the inclusion of the
results obtained with the CEDF PC-PK1 (limited so far to the isotopic 
chain with $Z=70$) shows that systematic uncertainties increase by a
factor of around 2.5 as compared with those presented in Fig.\
\ref{DE_spread_total}. Note that the addition of the PC-PK1 results
is not expected to alter much the spreads of binding energies
within the limit of nuclei reachable with FRIB \cite{AA.16}.

  It is interesting to compare our results of the analysis of statistical
uncertainties with Skyrme results based on the functional UNEDF0 
\cite{KENBGO.13,GDKTT.13}. While the statistical uncertainties are similar 
for binding energies in both approaches (compare Fig.\ 1 in Ref.\ 
\cite{GDKTT.13} with Fig.\ \ref{ground-stat-uncert}a in
the present paper), they are substantially smaller for the neutron skins in
relativistic results (compare Fig.\ 2 in Ref.\ \cite{KENBGO.13} with Fig.\
\ref{ground-stat-uncert}a in the present paper).

   The increase of statistical and systematic uncertainties on approaching
the neutron drip line clearly poses a challenge for theory. In CDFT, it
is dominated by the isovector channel of the effective interaction and its
density dependence. For example, the freezing of the coupling constant for the
$\rho$-meson in the functional NL3* during a selection of ``reasonable'' functionals 
leads to statistical uncertainties at the neutron drip line which are substantially 
smaller than those seen in Fig.\ \ref{ground-stat-uncert}. However, an improvement of
the isovector channel is not that simple. Two possible ways have been
considered in Ref.\ \cite{AA.16}. First, new mass measurements with
future rare isotope beam facilities will, in principle, improve isovector properties of
the CEDFs in the $Z\leq 50$ nuclei (where according to Fig.\ \ref{FRIB-impact}
most of the data will be measured). However, the improvement is expected to be modest.
Second, the improvement in nuclear matter properties will not substantially
reduce the uncertainties in the description of neutron-rich systems.

\section{Fission barriers in superheavy nuclei}
\label{FB}

\begin{figure*}[ht]
\centering
\includegraphics[angle=0,width=9cm]{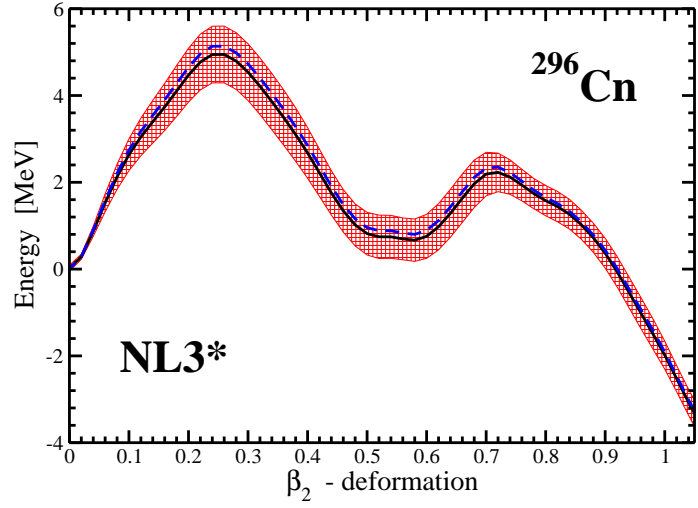}
\caption{Statistical uncertainties in the deformation energy
curves of the nucleus $^{296}$Cn. The mean potential energy curve
is shown by a solid black line. The red dashed region shows the
standard deviations in energy. The potential energy curve obtained
with the original functional NL3* is shown by a blue dashed line.}
\label{FB-stat-uncert}
\end{figure*}

 Another extreme of the nuclear landscape (high-$Z$ extreme) is the region of superheavy
elements (SHE). The structure of SHEs has recently been reexamined within CDFT  in Ref.\
\cite{AANR.15}. This led to significant revisions in our understanding of their
structure. Contrary to the previous CDFT studies, it was found that the impact of
the $N = 172$ spherical shell gap on the structure of SHEs is very limited. Similar
to non-relativistic functionals, some covariant functionals predict an important
role played by the spherical $N = 184$ gap. For these functionals (NL3*, DD-ME2, and
PC-PK1) there is a band of spherical nuclei along and near the $Z = 120$ and $N = 184$
lines. However, for other functionals (DD-PC1 and DD-ME$\delta$) oblate shapes dominate
at and in the vicinity of these lines. The available experimental data on SHEs are, in general,
described with comparable accuracy with these functionals. This makes it impossible
to discriminate between their predictions for nuclei outside the presently known region.

  The stability of SHEs is defined by the fission barriers. Thus, the study of systematic
and statistical uncertainties in the predictions of fission barriers has been undertaken
using the same of set CEDFs as in Sect.\ \ref{NR}.

\begin{figure*}[ht]
\centering
\includegraphics[angle=-90,width=11cm]{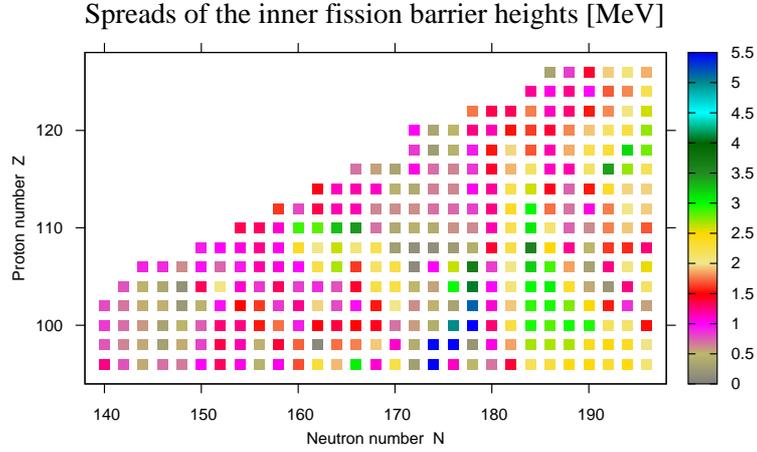}
\caption{The spreads $\Delta E^B$ of the heights of the inner fission 
barriers as a function of proton and neutron number.
$\Delta E^B(Z,N) = |E^{B}_{max}(Z,N)-E^{B}_{min}(Z,N)|$, where, for
given $Z$ and $N$ values, $E^{B}_{max}(Z,N)$ and $E^{B}_{min}(Z,N)$ are
the largest and smallest heights of inner fission barriers obtained
in axial RHB calculations with the set of functionals NL3*, DD-PC1,
and PC-PK1.}
\label{fission_spread}
\end{figure*}

  Statistical uncertainties in the deformation energy curves and fission barriers are
illustrated in Fig.\ \ref{FB-stat-uncert} on the example of the nucleus $^{296}$Cn. The
calculations are performed within the axial RHB framework with the functional NL3* \cite{NL3*}. 
Statistical uncertainties are quantified by the standard deviations in energy
$\sigma_E$ around the mean value of energy. These quantities are defined as a function of
deformation for a set of ``physically reasonable'' functionals using the formalism of
Refs.\ \cite{DNR.14,Stat-an}. They are small in the vicinity of the spherical minimum but
then they grow with increasing deformation. They become especially pronounced in the
vicinity of the inner and outer saddles and in the region of the superdeformed (SD)
minimum. Statistical uncertainties decrease substantially and stabilize above
the outer fission barrier. The calculations with the functional DD-ME2 lead to comparable
results but with a different deformation dependence of statistical uncertainties
\cite{AARR.17}.
Both calculations suggest that the increase of statistical uncertainties at some
deformation may be due to the underlying single-particle structure. This is because
the variations of the functional lead to modifications of the single-particle energies
as well as to changes in the sizes of the superdeformed shell gaps and the single-particle
level densities at the saddles and the SD minimum. These changes then affect the shell
correction energies.

\begin{figure}[ht]
\includegraphics[angle=0,width=6.2cm]{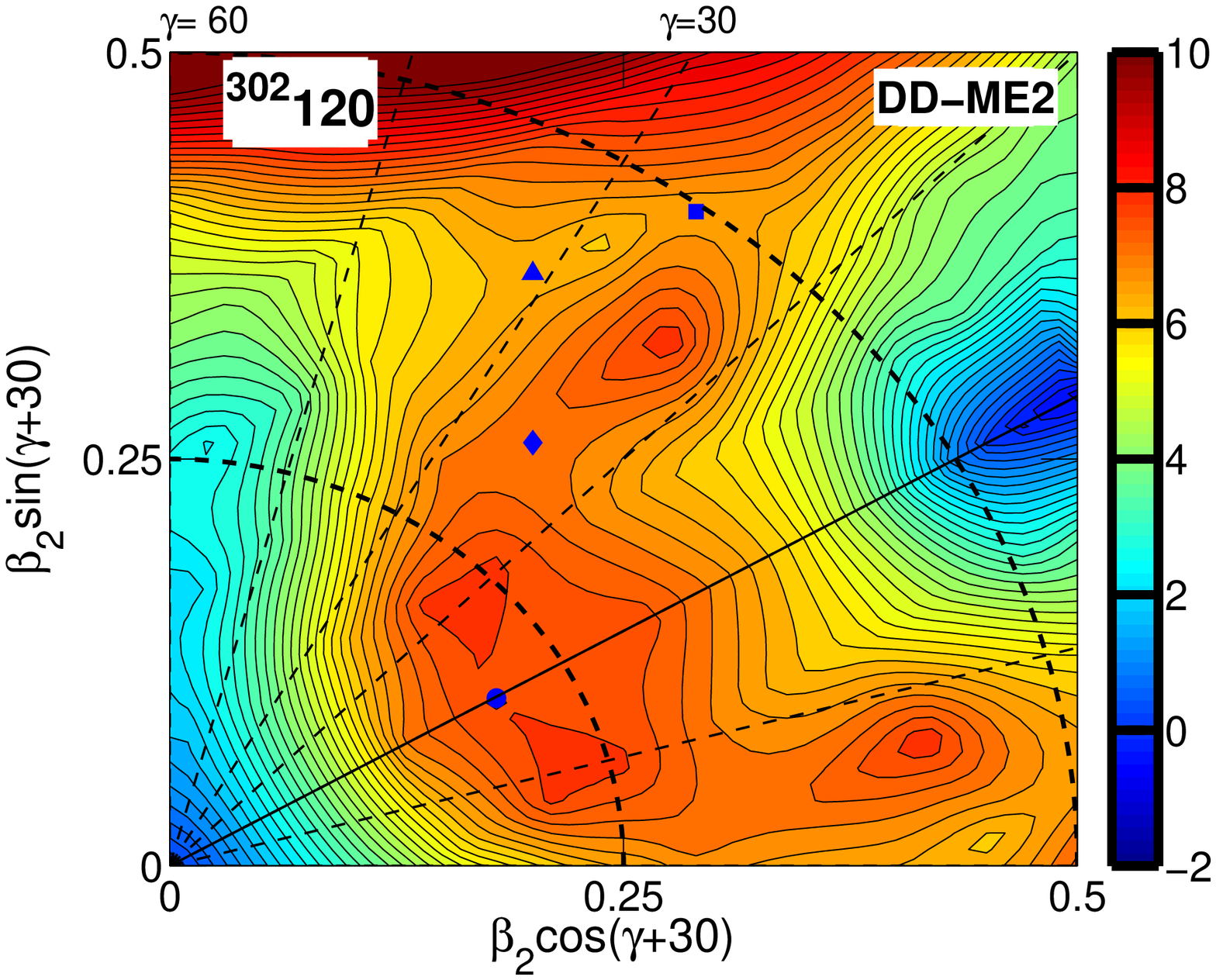}
\includegraphics[angle=0,width=6.2cm]{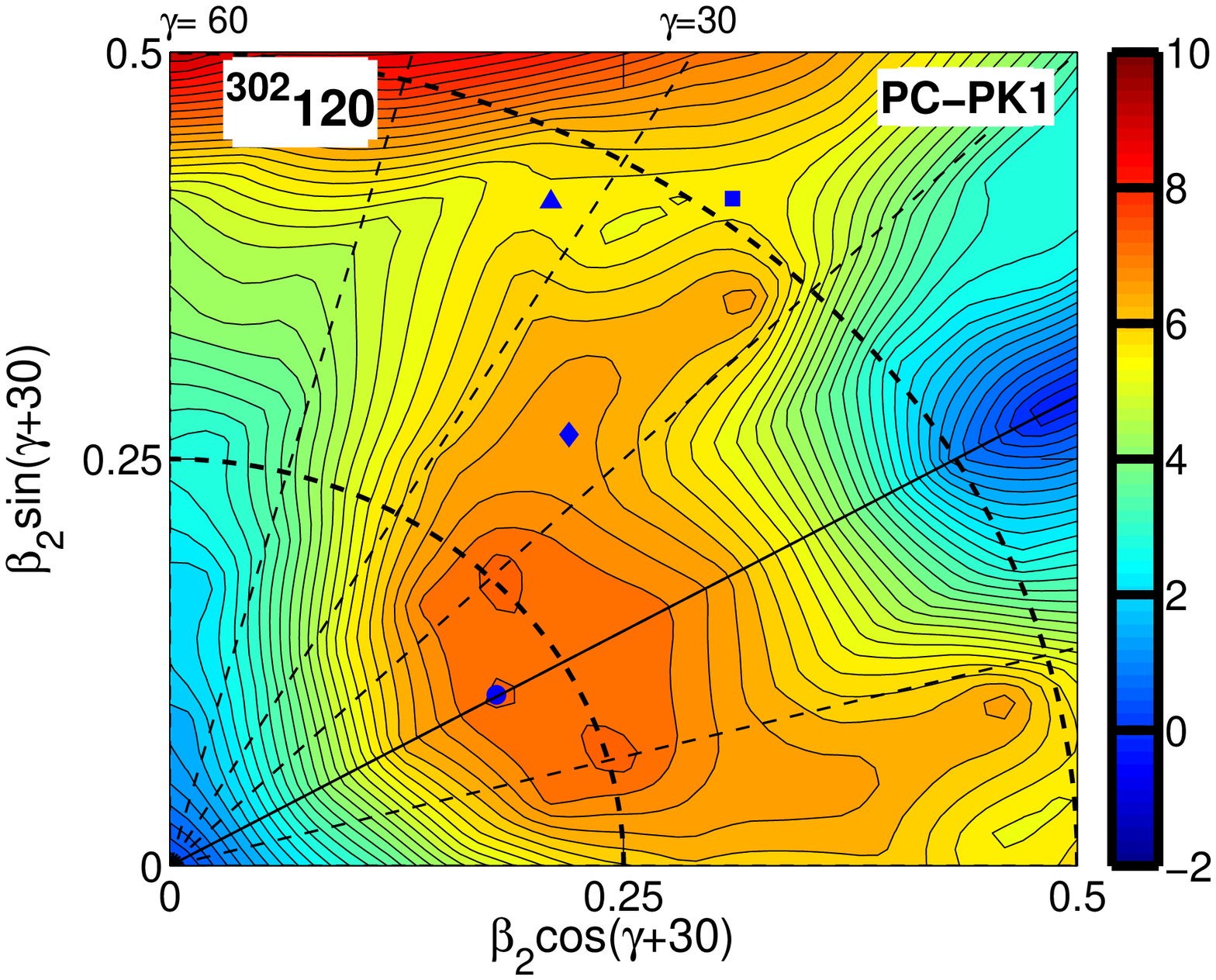}
\includegraphics[angle=0,width=6.2cm]{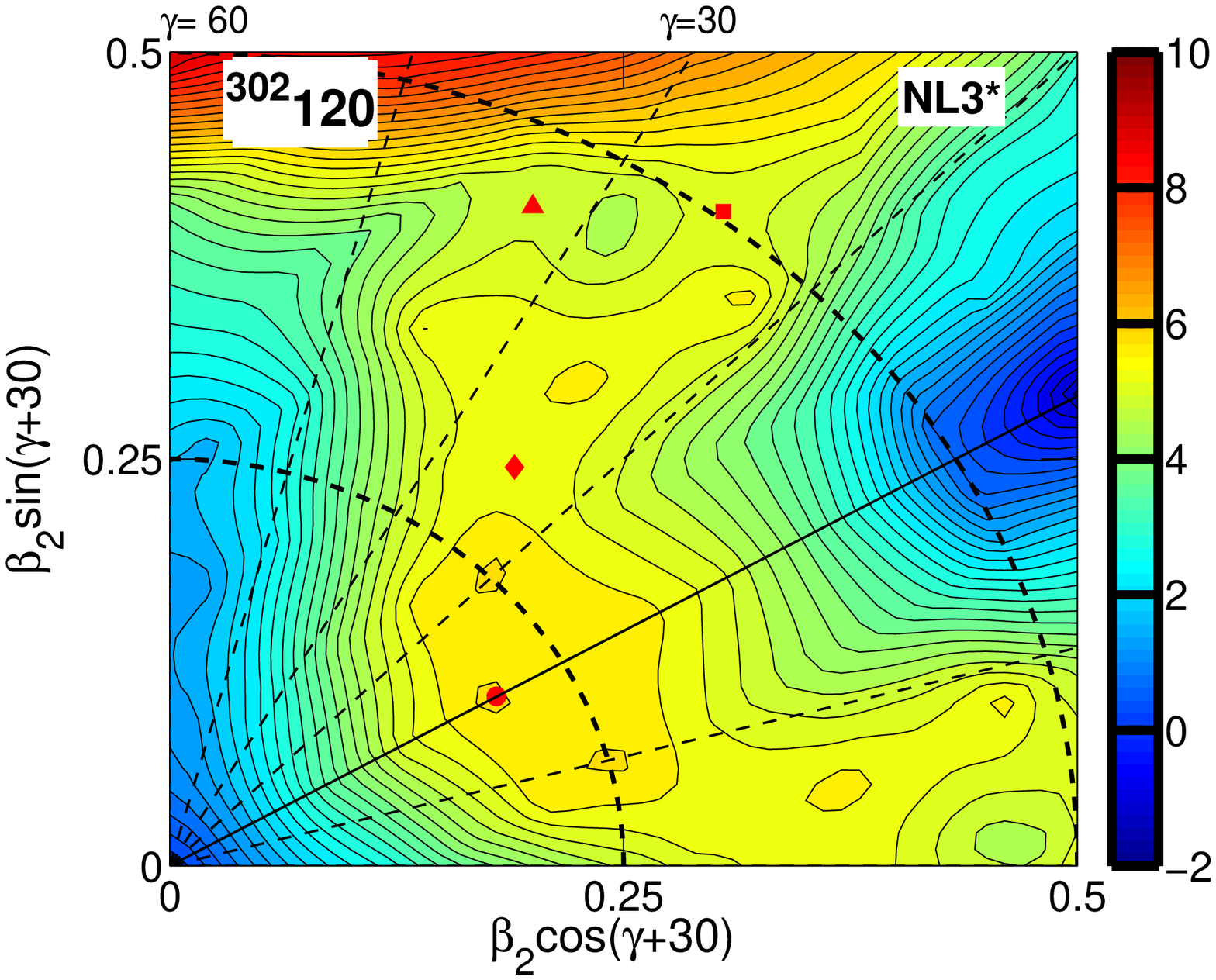}
\includegraphics[angle=0,width=6.2cm]{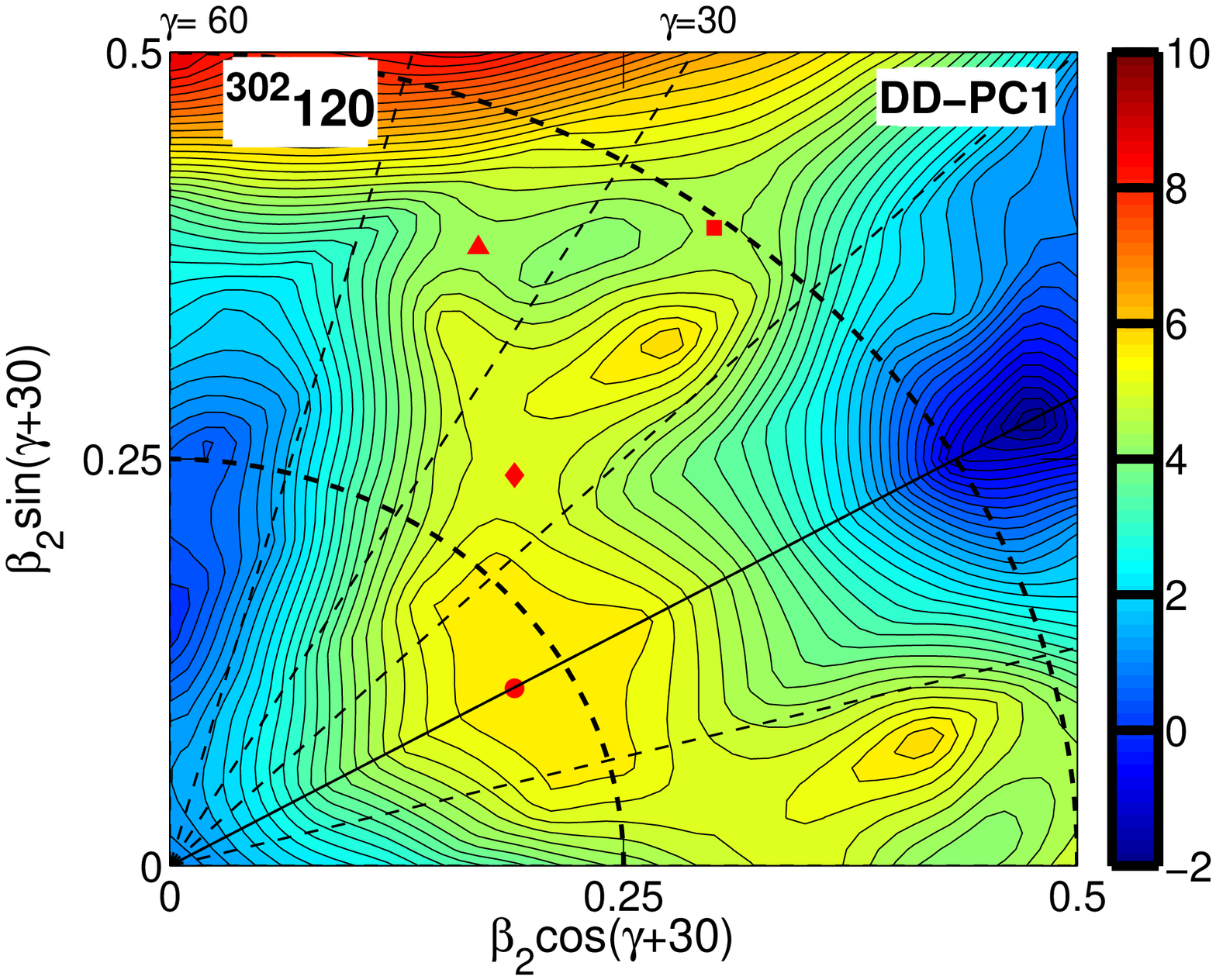}
\centerline{%
\includegraphics[angle=0,width=6.2cm]{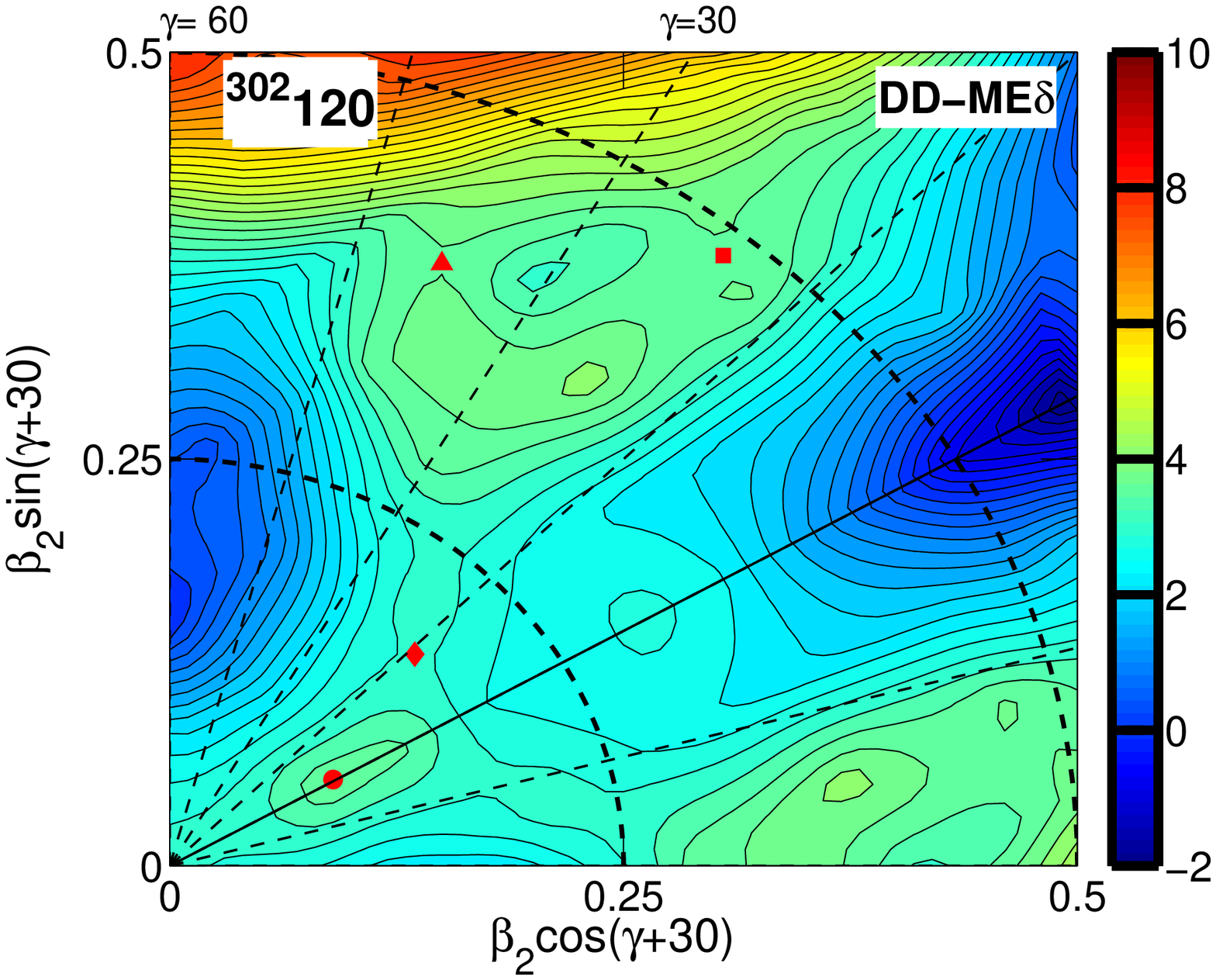}
}
\caption{Potential energy surfaces of the nucleus $^{302}120$ as
obtained in the calculations with the indicated CEDFs. The energy
difference between two neighboring equipotential lines is equal
to 0.25 MeV. The Ax, Ax-Tr, Tr-A and Tr-B saddles are shown by
blue/red circles, diamonds, triangles, and squares, respectively.
The PES are shown in the order of decreasing height of the inner
fission barrier.}
\label{Fig-pes-120}
\end{figure}

  The systematic uncertainties obtained in axially symmetric RHB
calculations for inner fission barrier heights are summarized in Fig.\
\ref{fission_spread}. The consideration here is restricted to three CEDFs,
namely, NL3*, DD-PC1 and PC-PK1. These functionals, fitted only to the
ground state properties of very limited set of nuclei (see details in
Ref.\ \cite{AARR.14}), successfully describe experimental fission barriers in
the actinides \cite{{AAR.10,AAR.12,LZZ.12,PNLV.12,LZZ.14}}. Theoretical
uncertainties, expressed in terms of the $\Delta E^B$ spreads, are typically
less than 2 MeV for the $N\leq 180$ nuclei; only in few nuclei around $Z=110,
N\sim 164$ and $Z\sim 110, N\sim 176$ these uncertainties are higher reaching
4 and 5.5 MeV respectively. However, these uncertainties increase by roughly
1 MeV for the nuclei with $N\geq 182$. It is also important to mention that
theoretical spreads in the inner fission barrier heights do not form a smooth
function of proton and neutron numbers; there is always random
component in their behavior.

  It is well known that inner fission barriers in many SHEs are affected by 
triaxiality; its impact is especially pronounced in the nuclei near the 
$Z=120$ and $N=184$ lines (Ref.\ \cite{AAR.12}). This is exemplified in 
(Fig.\ \ref{Fig-pes-120}) for the nucleus $^{302}$120 by potential energy 
surfaces (PESs). In this nucleus the triaxial saddles
(labeled as 'Tr-Ax', 'Tr-A', 'Tr-B') are located at $(\beta_2 \sim 0.32,
\gamma \sim 22^{\circ})$, $(\beta_2 \sim 0.43, \gamma \sim 34^{\circ})$, and
$(\beta_2 \sim 0.50, \gamma \sim 22^{\circ})$ for the functionals DD-ME2,
PCPK1, NL3* and DD-PC1. The 'Tr-A' and 'Tr-B' saddles are present also
in the PES for DD-ME$\delta$, but for this functional the 'Tr-Ax' saddle
is shifted to smaller $\beta_2$ and $\gamma$ deformations. The axial
saddle is higher in energy than the lowest in energy triaxial saddle
for all functionals. Note that the topology of the PES for the functional 
DD-ME$\delta$ differs substantially from the one for other functionals.

\begin{figure*}[ht]
\centering
\includegraphics[angle=-90,width=6.2cm]{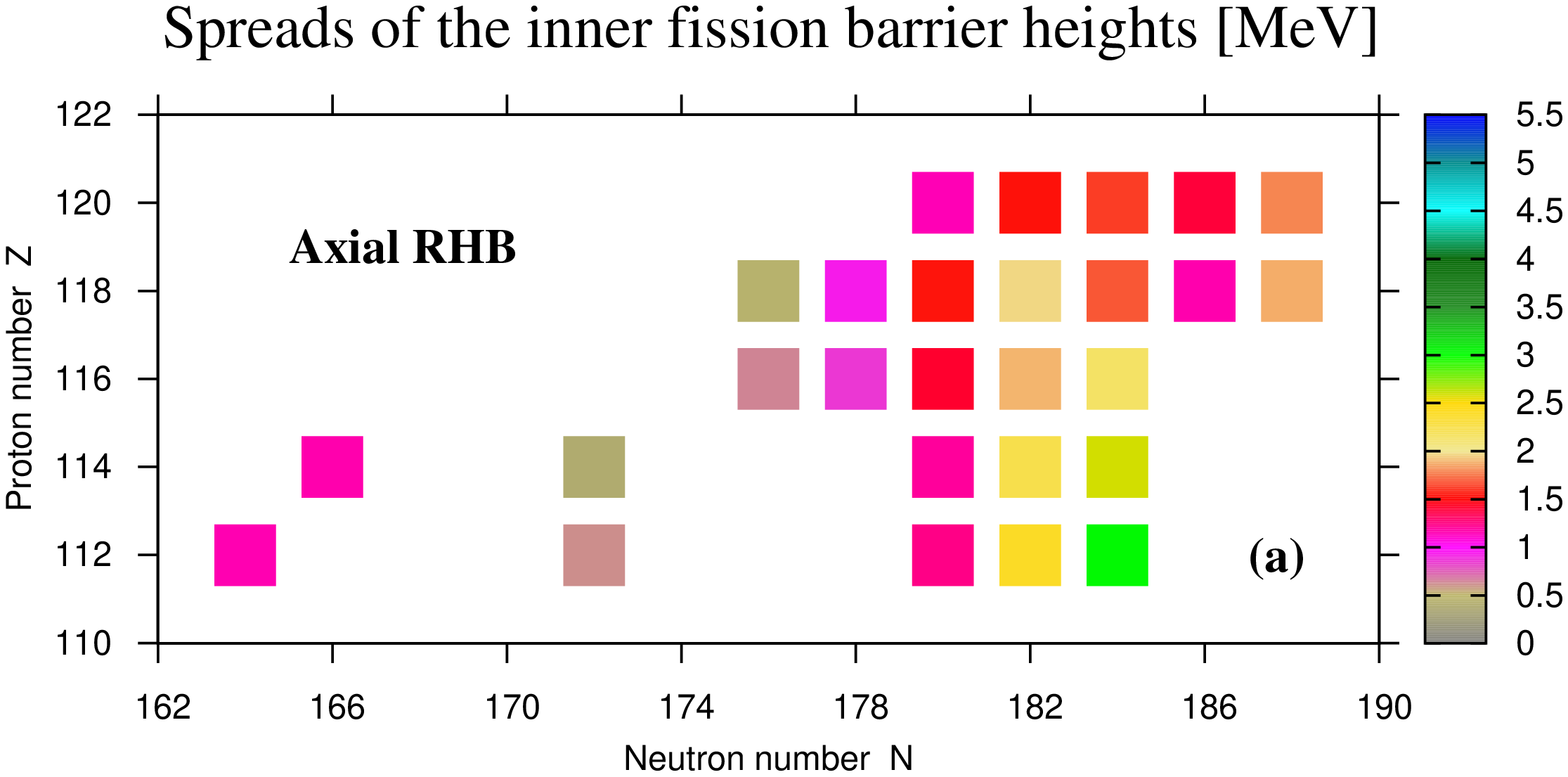}
\includegraphics[angle=-90,width=6.2cm]{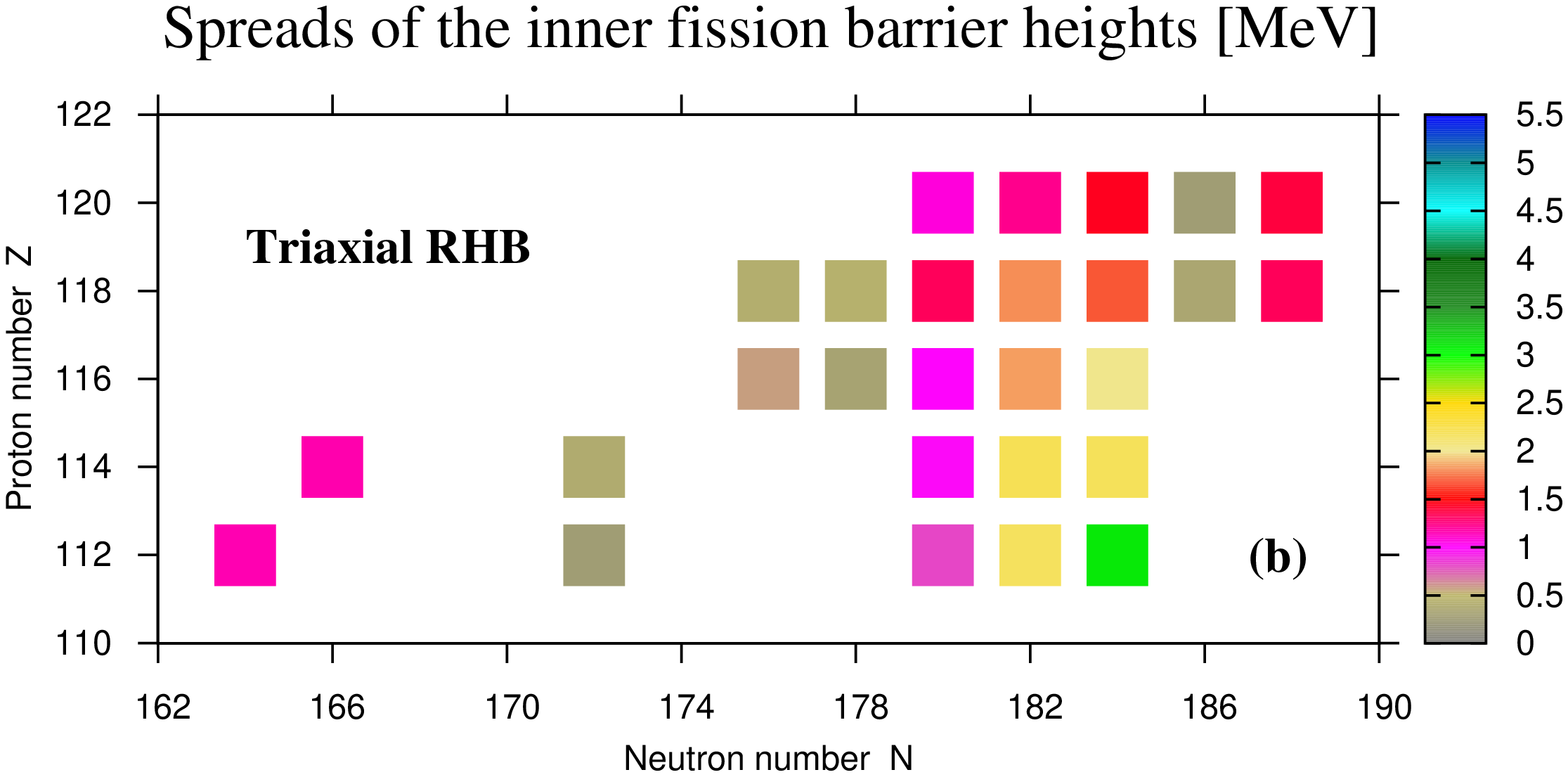}
\caption{The same as in Fig.\ \ref{fission_spread} but for a selected set
of the $Z=112-120$ nuclei. Panels (a) and (b) show the spreads $\Delta E^B$
obtained in axial and triaxial RHB calculations, respectively.}
\label{FB-stat-uncert-triax}
\end{figure*}

  The accounting of triaxiality in the calculations modifies the spreads
in the predictions of the heights of inner fission barriers. This is
clearly seen in Fig.\ \ref{FB-stat-uncert-triax} where these spreads,
obtained in axial and triaxial RHB calculations, are compared. Although,
locally, two calculations may differ slightly, on average there are strong
correlations in the spreads obtained in the two calculations. This suggests
that also for other regions of the nuclear chart, not covered by the present
triaxial RHB calculations, the spreads in inner fission barrier heights
obtained in the axial RHB calculations (see Fig.\ \ref{fission_spread})
could be used as a reasonable estimate of the spreads which would be
obtained in the calculations with triaxiality included.

  The benchmarking of the functionals to experimentally known fission
barriers in the actinides reduces the number of suitable functionals
to three (NL3*, DD-PC1 and PC-PK1). This allows to decrease theoretical
uncertainties in inner fission barrier heights since the $\Delta E^{B}$
spreads for five functionals are substantially higher than those
presented  in Figs.\ \ref{fission_spread} and \ref{FB-stat-uncert-triax}
(see also the discussion in Ref.\ \cite{AARR.17}). This fact is clearly
seen also in Fig.\ \ref{Fig-pes-120}. Even those reduced uncertainties
of the inner fission barrier heights translate into the uncertainties of
many orders of magnitude for spontaneous fission half-lives (see Ref.\
\cite{AARR.17}).

\section{Conclusions}
\label{concl}

   In order to quantify theoretical predictions for covariant density 
functional theory in unknown regions of the nuclear chart, systematic 
uncertainties deduced from the results of a set of different well known
covariant energy density functionals, as well as statistical uncertainties 
derived according to Ref.\ \cite{DNR.14} for a set of ``reasonable'' 
variations of one functional, are discussed for ground state observables
such as binding energies and neutron skin thicknesses over entire nuclear
chart and for inner fission barriers in superheavy nuclei. It is found 
that the statistical uncertainties are usually smaller than the systematic 
ones. We observe a systematic growth of the uncertainties 
for increasing deviations from the experimentally known regions, in 
particular, when approaching the neutron drip line or the region of
superheavy nuclei with extreme $Z$ values.
 
  Of course, the present investigations are restricted to the mean field level. 
Employed covariant energy density functionals are fitted to nuclear matter properties 
and to ground state properties  of finite nuclei, such as binding 
energies and charge radii. Therefore, one can expect that in the 
future, when we are able to take into account the beyond mean field 
effects in a microscopic way not only at the model level but also 
in the fitting protocols, the predictive power of CDFT will increase
considerably with appropriate reduction in systematic and statistical
uncertainties.

%

\section{Acknowlegdments}
 This material is based upon work supported by the Department of
Energy National Nuclear Security Administration under Award Number
DE-NA0002925, by the U.S. Department of Energy, Office of Science,
Office of Nuclear Physics under Award Number DE-SC0013037 and by
the DFG cluster of excellence \textquotedblleft Origin and Structure
of the Universe\textquotedblright\
(www.universe-cluster.de).


\begin{thebibliography}{99}

\bibitem{MSMA.16} M.~ R.~ Mumpower and R.~ Surman and G.~ C.~ McLaughlin and
                   A.~ Aprahamian, Prog. Part. Nucl. Phys. 86, 0146 (2016)

\bibitem{GBJ.11} S.\ Goriely and A.\ Bauswein and H.-T.~ Janka,
                 Astr.\ Phys. J 738, L32 (2011).

\bibitem{DNR.14} J.~ Dobaczewski and W.~ Nazarewicz and P.-G.~ Reinhard,
  J.~ Phys.~ G 41, 074001 (2014)

\bibitem{VALR.05} D.~ Vretenar and A.~ V.~ Afanasjev and G.~ A.~ Lalazissis
        and P.~ Ring, Phys. Rep. 409, 101 (2005).

\bibitem{TW.99} S.\ Typel and H.\ H.\ Wolter, Nucl. Phys. A656, 331 (1999).

\bibitem{DD-ME2} G.~ A.~ Lalazissis and T.~ Nik{\v{s}}i{\'{c}} and
                  D.~ Vretenar and P.~ Ring, Phys. Rev. C 71, 024312 (2005).

\bibitem{DD-PC1} T.~ Nik\v{s}i\'{c} and D.~ Vretenar and P.~ Ring, Phys. Rev. C,
                 78, 034318 (2008).

\bibitem{BB.77} J.\ Boguta and R.\ Bodmer, Nucl.\ Phys. A292, 413 (1977).

\bibitem{NL3*} G.~ A.~ Lalazissis and S.~ Karatzikos and R.~ Fossion and
               D.~ Pe{\~n}a~Arteaga and A.~ V.~ Afanasjev and P.~ Ring,
               Phys.\ Lett. B671, 36 (2009).

\bibitem{DD-MEd}  X.\ Roca-Maza and X.\ Vi{\~n}as and M.\ Centelles
                 and  P.\ Ring and P.\ Schuck, Phys. Rev. C 84, 054309 (2011).

\bibitem{PC-PK1} P.~ W.~ Zhao and Z.~ P.~ Li and J.~ M.~ Yao and J.~ Meng,
         Phys. Rev. C 82, 054319 (2010).



\bibitem{AARR.14} S.~ E.~ Agbemava and A.~ V.~ Afanasjev and D.~ Ray and P.~ Ring,
               Phys.\ Rev. C 89, 054320 (2014).

\bibitem{AARR.13} A.~ V.~ Afanasjev and S.~E.~ Agbemava and D.~ Ray and P.~ Ring,
         Phys. Lett. B 726, 680 (2013).

\bibitem{AA.16} A.~ V.~ Afanasjev and S.~ E.~ Agbemava,
                Phys. Rev. C 93, 054310 (2016).

\bibitem{AAR.16} S.~ E.~ Agbemava and A.~ V.~ Afanasjev and P.~ Ring,
         Phys. Rev. C 93, 044304 (2016).

\bibitem{AANR.15} S.\ E.\ Agbemava, A.\ V.\ Afanasjev, T.\ Nakatsukasa, and 
                  P.\ Ring, Phys. Rev. C 92, 054310 (2015).

\bibitem{Eet.12} J.~ Erler and N.~ Birge and M.~ Kortelainen and
             W.~ Nazarewicz and E.~ Olsen and A.~ M.~ Perhac and
             M.~ Stoitsov, Nature 486, 509 (2012).

\bibitem{KENBGO.13} M.~ Kortelainen and J.~ Erler and W.~ Nazarewicz and
        N.~ Birge and Y.~ Gao and E.~ Olsen, Phys. Rev. C 88, 031305(R),
        (2013).

\bibitem{GDKTT.13} Y.\ Gao and J.\ Dobaczewski and M.\ Kortelainen and
        J.\ Toivanen and D.\ Tarpanov, Phys. Rev. C 87, 034324 (2013).

\bibitem{MSHSWN.15} J.\ D.\ McDonnell, N.\ Schunck, D.\ Higdon, J.\ Sarich,
            and S.\ M.\ Wild and W. Nazarewicz, Phys. Rev. Lett. 114, 122501
            (2015).

\bibitem{Stat-an} S.~ Brandt, {\it Data analysis. Statistical and Computational
           Methods for Scientists and Engineers}, (Springer International Publishing,
           Switzerland, 2014)

\bibitem{AARR.17} S.\ E.\ Agbemava, A.\ V.\ Afanasjev,  D.\ Ray and P. Ring,
                  submitted to Phys. Rev. C

\bibitem{AAR.10} H.~ Abusara and A.~ V.~ Afanasjev and P.~ Ring,
Phys.\ Rev. C 82, 044303 (2010).

\bibitem{AAR.12} H.~ Abusara and A.~ V.~ Afanasjev and P.~ Ring,
Phys.\ Rev. C 85, 024314 (2012).

\bibitem{LZZ.12} B.-N.\ Lu, E.-G.\ Zhao and S.-G.\ Zhou, Phys.\
Rev. C 85, 011301 (2012).

\bibitem{PNLV.12} V.\ Prassa, and T.\ Nik\ifmmode \check{s}\else \v{s}\fi{}i\ifmmode \acute{c}\else \'{c}\fi{},
         G.\ A.\ Lalazissis and D.\ Vretenar, Phys. Rev. C 86, 024317 (2012)

\bibitem{LZZ.14} B.-N.\ Lu, J.\ Zhao, E.-G.\ Zhao, S.-G.\ Zhou, Phys. Rev.
C 89, 014323 (2014).
















\end{thebibliography}

\end{document}